\documentclass[sigconf]{acmart}

\usepackage{listings} 
\usepackage{xcolor}
\usepackage{mdframed} 
\definecolor{codered}{rgb}{0.6,0,0}
\definecolor{codegreen}{rgb}{0,0.6,0}
\definecolor{codegray}{rgb}{0.5,0.5,0.5}
\definecolor{codepurple}{rgb}{0.58,0,0.82}
\definecolor{backcolour}{rgb}{0.95,0.95,0.92}
\lstdefinestyle{mystyle}{
    backgroundcolor=\color{backcolour},   
    commentstyle=\color{codegreen},
    keywordstyle=\color{magenta},
    numberstyle=\tiny\color{codegray},
    stringstyle=\color{codepurple},
    basicstyle=\ttfamily\footnotesize,
    breakatwhitespace=false,         
    breaklines=true,                 
    captionpos=b,                    
    keepspaces=true,                 
    numbersep=5pt,                  
    showspaces=false,                
    showstringspaces=false,
    showtabs=false,                  
    tabsize=2
}
\usepackage{enumitem}
\usepackage{xcolor}
\usepackage{tikz}
\definecolor{customMagenta}{HTML}{E52CCF}

\AtBeginDocument{%
  \providecommand\BibTeX{{%
    \normalfont B\kern-0.45em{\scshape i\kern-0.25em b}\kern-0.8em\TeX}}}

\copyrightyear{2026}
\acmYear{2026}
\setcopyright{cc}
\setcctype{by}
\acmConference[ITiCSE 2026]{Proceedings of the 31st ACM Conference on Innovation and Technology in Computer Science Education V. 1}{July 10--15, 2026}{Madrid, Spain}
\acmBooktitle{Proceedings of the 31st ACM Conference on Innovation and Technology in Computer Science Education V. 1 (ITiCSE 2026), July 10--15, 2026, Madrid, Spain}
\acmDOI{10.1145/3803400.3809386}
\acmISBN{979-8-4007-2634-7/2026/07}

\begin{document}


\title[Neurodivergent Computing Students' Experiences with Collaborative Active Learning]{``I can't read your mind'': A Study of Neurodivergent Computing Students' Experiences with Collaborative Active Learning}


\author{Cynthia Zastudil}
\affiliation{%
\institution{Temple University}
\city{Philadelphia}
\state{PA}
\country{USA}}
\email{cynthia.zastudil@temple.edu}

\author{Srishty Muthusekaran}
\affiliation{%
\institution{Temple University}
\city{Philadelphia}
\state{PA}
\country{USA}}
\email{srishty.muthusekaran@temple.edu}

\author{Rayhona Nasimova}
\affiliation{%
\institution{Temple University}
\city{Philadelphia}
\state{PA}
\country{USA}}
\email{rayhana.nasimova@temple.edu}

\author{Stephen MacNeil}
\affiliation{%
\institution{Temple University}
\city{Philadelphia}
\state{PA}
\country{USA}}
\email{stephen.macneil@temple.edu}


\renewcommand{\shortauthors}{Cynthia Zastudil, Srishty Muthusekaran, Rayhona Nasimova, \& Stephen MacNeil}


\begin{abstract}
Computing courses often feature active learning techniques that promote collaboration and social interaction between students. However, neurodivergent students' preferences and experiences with these techniques are not well understood. We conducted a survey of neurodivergent computing students (n=24), specifically autistic students or students with ADHD, and neurotypical computing students (n=20) to understand how the structure of collaborative active learning affects their comfort in computing courses. We also interviewed four computing students on the autism spectrum or with ADHD to gain more contextualized insights into their experiences and accessibility recommendations. Our survey surfaces how team dynamics and assignment structure can impact neurodivergent students' comfort in computing courses. Neurodivergent students expressed discomfort with assignments that lack structure or have ambiguous expectations. Neurodivergent students prefer smaller teams that work together frequently with explicitly defined roles. Our interviews identified ways that neurodivergent students cope with discomfort in collaborative active learning, including self-selecting roles and self-disclosure. While preliminary, our results highlight how instructors can design collaborative active learning to be more equitable and accessible for neurodivergent students.
\end{abstract}



\maketitle

\section{Introduction}
Active learning has become a popular pedagogical approach in computing education~\cite{sanders2017folk, papamitsiou2020computing}. Active learning techniques encourage students to engage more deeply with course material through doing and self-reflection~\cite{bonwell1991active}.
Collaborative active learning (CAL) requires students to work together in and out of class on activities~\cite{sanders2017folk}.
While these features may benefit many students, they could introduce barriers for neurodivergent (ND) students, such as autistic students or students with attention deficit hyperactivity disorder (ADHD)~\cite{pilotte2016autism,stuurman2019autism,sharmin2024towards}\footnote{We use both identity-first language (e.g., ``autistic students'') and person-first language (e.g., ``students with ADHD''). People with ADHD often prefer person-first language, while the preferences of autistic people vary.~\cite{nah2025preferences, sharif2022should}},
who may face challenges with social interaction, executive function, or processing abstract instructions~\cite{clouder2020neurodiversity, zolyomi2018values}. 

Computing education research has not historically focused on the experiences of ND students~\cite{luchs2021considering, koushik2019broadens, zastudil2025neurodiversity}. 
The specific learning strengths and challenges of ND students are not well understood, especially in the context of computing education~\cite{borsotti2024neurodiversity}.
We conducted a survey with neurotypical (NT) and ND computing students to understand how aspects of CAL might impact NT and ND students, specifically autistic students or students with ADHD\footnote{For brevity, we use ``neurodivergent'' to refer to autistic students or students with ADHD in this paper.}, differently.
We investigated the following research questions:
\begin{itemize}
    \item[\textbf{RQ1:}] How do aspects of team dynamics impact the comfort of ND computing students with CAL?
    \item[\textbf{RQ2:}] How do pedagogical and environmental aspects (e.g., instructor facilitation, assignment guidelines, and location) impact the comfort of ND computing students with CAL?
    \item[\textbf{RQ3:}] In what ways do ND computing students adapt to be more comfortable with CAL and what recommendations do they have to make it more accessible?
\end{itemize}

We collected 44 survey responses from ND (n=24) and NT (n=20) computing students and conducted 4 interviews with ND computing students.
We chose to include both NT and ND students in our survey because we wanted to understand what aspects of CAL disproportionately affect ND students' comfort in their computing courses.
Our survey surfaced how decisions about CAL structure can impact how comfortable ND students feel in computing courses. Specifically, ND students prefer smaller, recurring teams with explicitly defined roles. ND students felt uncomfortable with assignments that lack structure or have ambiguous expectations. Our interviews identified ways that ND students deal with discomfort and adapt their behavior to accommodate to CAL, including self-selecting roles and self-disclosure. 
While preliminary, our findings can aid instructors
in developing more equitable CAL techniques.

\section{Related Work}
The experiences of ND students have not been widely studied across computing education. A recent literature review found only 14 peer-reviewed research papers published on the topic~\cite{zastudil2025neurodiversity}. Studies in this space also acknowledge this issue, and call for more attention to the area~\cite{hirst2025neurodivergence}. Other studies on neurodiversity in higher education, especially in STEM programs, further make the point that this is a wide-spanning gap across higher education~\cite{bonnette2025need, koushik2019broadens}.

Autistic students and students with ADHD are more likely to enroll in computing-related degree programs~\cite{shifrer2021problematizing, hirst2025neurodivergence, mettenbrink2025prevalence}. These neurodiverse identities are more likely to co-occur and can experience similar challenges with executive function (e.g., time management, planning, and self-regulation), socialization, and hyperactivity~\cite{antshel2019autism, craig2015overlap}.
Prior research indicates that it is important to recognize both how these overlaps affect ND students and how instructors should view making their courses more inclusive~\cite{bonnette2025need,borsotti2024neurodiversity}.

CAL techniques, however, have been widely studied across computing education. The computing education community generally views active learning methods as a positive addition to courses~\cite{sanders2017folk}, as there is evidence that these techniques can improve student success, sense of belonging, and retention~\cite{moudgalya2021measuring, latulipe2018evolving, michael2006where, greer2019effects, latulipe2025investigating}. 
The introduction of CAL has shown improvement in sense of belonging for marginalized groups~\cite{runa2023student, tadimalla2025connecting, latulipe2025investigating}. However, the impact it has on ND students is not as comprehensively understood. Despite the breadth of research on CAL techniques conducted within STEM programs, many studies do not highlight that these students may be having different experiences than their NT peers. 

Concern has been raised over how educators often lack the knowledge and resources needed to make the pedagogical transition to a classroom that implements CAL techniques with consideration of its ND students~\cite{pilotte2016autism}. A study with ND STEM students found that group work can be a source of anxiety and difficulty~\cite{salvatore2024not}. However, this study did not include many computing students. Computing education employs unique CAL techniques~\cite{sanders2017folk}, like pair programming, and understanding the experiences of ND computing students is important to ensure equitable pedagogical design. We hope to uncover both the effects of CAL on ND students and how instructors can better facilitate the learning of ND students in collaborative settings through more informed teaching practices~\cite{pilotte2016autism, salvatore2024not}. 

\section{Methodology}
We conducted both survey and interview studies to investigate our research questions. All research was conducted after receiving approval from our university's Institutional Review Board. Our goal in these studies was not to reach saturation or achieve generalizability; rather, we wanted to begin the process of understanding how ND students' comfort and experiences are impacted by CAL.

\subsection{Participant Recruitment}
We recruited from multiple North American universities for both of our research methods. We conducted convenience sampling by advertising in computing-related student Discord servers and Slack channels. We also requested that instructional faculty and student organization leaders share our recruitment information with their students. Survey participants were not compensated and interview participants received a \$10 gift card. All data was collected between Aug.--Dec. 2025. 

\subsection{Survey Study}
Given that students' experiences with CAL vary widely based on team dynamics, instructors' pedagogical decisions, and environmental factors~\cite{monson2019they,shekhar2020negative,gordy2020multi}, we wanted to understand how the structure of CAL impacts the comfort of ND students.

\subsubsection{Participant Demographics}
Participants were asked to provide their gender, ethnicity, university, major, and their ND identity (if applicable).\footnote{Participants were able to use their preferred language for their gender and ethnicity, allowing them to provide us information going beyond a gender binary and narrow ethnic categories~\cite{mcgill2023conducting, oleson2022decade}. We collect and report this data to ensure that not only the dominant groups within computing were included in our data collection~\cite{mcgill2023conducting}.}
We collected 44 survey responses (27 men, 1 transgender man, 10 women, 1 transgender woman, 4 non-binary, and 1 genderfluid) from computing students from 12 North American universities. Participants' ethnicities included white (n=19), South Asian (e.g., Indian, Nepalese) (n=4), East Asian (e.g., Korean, Japanese) (n=4), Asian (n=9), African (e.g., Egyptian, Nigerian) (n=2), black (n=1), mixed (n=1), Middle Eastern (n=1),  and Hispanic and Latino (n=1), with 2 participants preferring not to respond. 

Participants included students who self-identified as autistic (n=5), having ADHD (n=12), and both autistic and having ADHD (n=7). 20 participants did not identify as ND. Similar to prior studies with ND participants~\cite{kirdani2024neurodivergent, bonnette2025need, runa2023student}, we did not require participants to have an official diagnosis to support their ND identity.
The process to receive a formal diagnosis for autism and ADHD is fraught with bias~\cite{seers2021you, begeer2009underdiagnosis}. If someone does not match the dominant identity of an ``autistic person'' or ``someone with ADHD'' they are less likely to receive a diagnosis~\cite{seers2021you, begeer2009underdiagnosis}. Social stigma, cultural influences, and lack of access to clinical evaluation may inform someone's decision to pursue a formal diagnosis and there are likely more people identifying as ND than are officially diagnosed~\cite{norbury2013difference, price2022unmasking}. 
We relied on participants' self-disclosed ND identities without requiring an official diagnosis. We focused on computing students on the autism spectrum and with ADHD to keep our work within a reasonable scope and not to homogenize the experiences of ``all'' ND students. 

\subsubsection{Design of Survey Questions}

Students are often unfamiliar with pedagogical terms such as ``peer instruction'' and ``think-pair-share'' because instructors often use more colloquial terms or forego naming activity types in class. Therefore, we asked participants more generally about their comfort levels with specific aspects of CAL. We structured our survey this way to understand how CAL techniques impact ND computing students' comfort in their courses independent of specific pedagogical techniques. We were interested in understanding participants' comfort levels with a wide variety of CAL activities, including in-class and longer-term activities (e.g., pair programming, think-pair-share, and semester-long group capstone projects).
We included the following aspects of CAL: \textit{team formation, team sizes, length of time working in a team, role assignment, levels of instructor involvement, assignment structure, and online versus in-person teamwork}. Participants were asked to rate their comfort on a seven-point Likert-scale from ``Very Uncomfortable'' to ``Very Comfortable.'' We asked participants about their comfort levels, because it is an important indicator of sense of belonging, student success, and retention~\cite{moudgalya2021measuring}.

\subsubsection{Quantitative Analysis}
We calculated the distribution of participants' responses to our survey questions and observed how participants' answers aligned and misaligned between NT and ND participants by comparing the distribution of participants' responses.

\subsection{Interview Study}
Our survey study investigated students' comfort with different aspects of CAL, but did not capture the details of the experiences of ND students which have impacted their comfort in their computing courses. We designed our interviews to engage directly with ND students to gain a deeper understanding of these experiences.

\subsubsection{Participants}
We conducted semi-structured interviews with 4 computing students (3 autistic students and 1 student who identified as both autistic and having ADHD) from 3 North American universities. As with our survey, participants were not required to have an official diagnosis of autism or ADHD. Interviews lasted between 35-45 minutes and were conducted remotely via Zoom.
Survey participation was not required for interview participation.

\subsubsection{Design of Interview Questions}
We conducted a semi-structured interview to be able to explore neurodivergent students' experiences with CAL. The semi-structured nature of our interview allowed us to ask follow-up questions as needed in order to better understand participants' preferences and experiences. 
We designed our interview questions to gain a better understanding of the underlying reasons for neurodivergent students' comfort with aspects of CAL.
We also asked participants to share strategies they developed to adapt to CAL and their ideas for making CAL more accessible (e.g., ``What strategies do you use to make CAL work better for you?'' and ``How do instructors/peers make CAL more/less accessible?'').

\subsubsection{Qualitative Analysis}

We conducted reflexive thematic analysis (RTA) of the interviews based on Braun and Clarke's guidelines~\cite{braun2019reflecting} to uncover insights not identifiable via the quantitative data collection from our survey. Two researchers first familiarized themselves with the transcripts from the interviews, then individually performed open coding~\cite{coding2004anselm} on the transcripts to identify initial codes. Once there was an initial set of codes, two researchers determined the final set of codes and grouped the codes into themes through periodic group discussions to refine their understanding of the data. In these discussions, researchers shared their interpretations of the data, and used any differing interpretations of the data to critically evaluate their own interpretations.
Our goal was not to reach a consensus, but to thoughtfully engage with the data~\cite{braun2019reflecting}.

\section{Positionality Statement}
Our team contains a mix of junior and senior researchers, including undergraduate students. All researchers have participated in CAL, and one member of our research team has taught courses with CAL.
We approach this work as allies to the ND student community. Our goal is not to dissuade the use of CAL, but rather to inform researchers and practitioners about ways in which CAL can be designed more equitably for more students.

\section{Results}
In Sections~\ref{sec:team-dynamics} and ~\ref{sec:other-impacts}, we present the results of our survey with relevant participant insights from our interviews. In Section~\ref{sec:interview-results}, we present findings from our interviews with ND students about how they adapt to succeed in CAL and their recommendations for making it more accessible for ND students. 

\subsection{RQ1: Team Dynamics}
\label{sec:team-dynamics}

\begin{figure*}
    \centering
    \includegraphics[width=\linewidth]{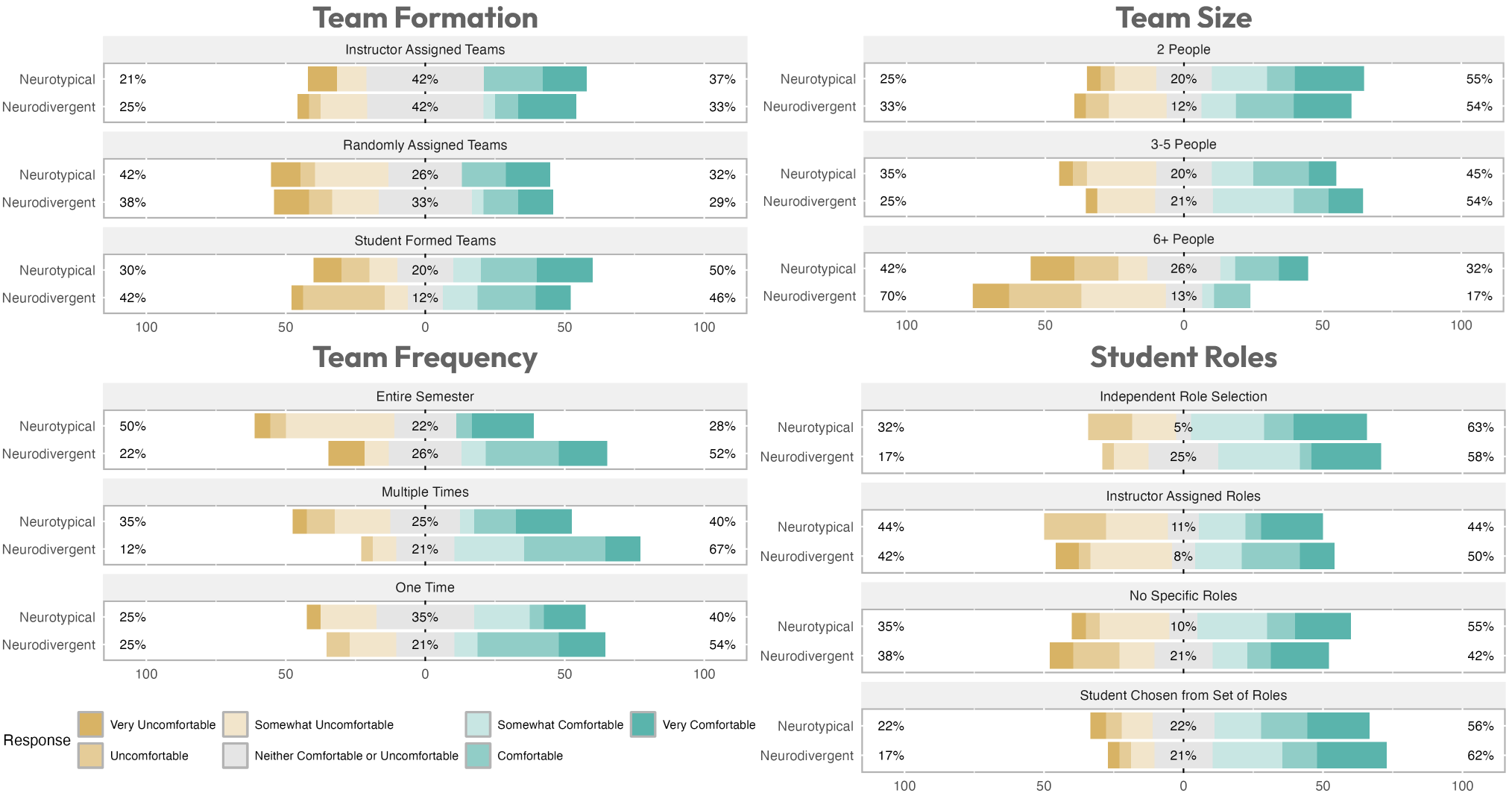}
    \caption{Percentage distributions of participants' responses to our Likert-scale survey questions on team dynamics.}
    \Description[A summary of Likert-scale survey question results focused on team dynamics.]{This figure displays four team dynamics topics from our survey: team formation, team size, team frequency, and student roles. Participants' responded on a seven-point Likert-scale from Very Uncomfortable to Very Comfortable for each question. Neurodivergent students preferred teams to be formed by instructors, then students, then randomly. Neurotypical students preferred teams to be formed by students, then instructors, then randomly. Neurodivergent students preferred teams to be 3-5 people, then 2 people, then 6 or more people. Neurotypical students preferred for teams to be 2 people, then 3-5 people, then 6 or more people. Neurodivergent students preferred for teams to meet multiple times, then last for the entire semester, then work together only one time. Neurotypical students preferred teams to work together one time, then multiple times, then for the entire semester. Neurodivergent students preferred explicit roles either independent chosen by students or from a instructor-defined set of roles, then instructor-assigned roles, then no explicit roles. Neurotypical students preferred student-selected roles either independently or from an instructor-defined set, then no explicit roles, then instructor-assigned roles.}
    \label{fig:dynamics}
\end{figure*}
 
Positive team dynamics play an important role in students' comfort in CAL and often result in better student outcomes~\cite{monson2019they}. 
Participants' responses to these survey questions are shown in Figure~\ref{fig:dynamics}.

\subsubsection{Team Formation}
\label{sec:formation}
ND students felt the most comfortable with student-formed teams
, then instructor-assigned teams, and lastly randomly-assigned teams. 
This order of preference was the same for our NT participants.
In our interviews, participants had conflicting opinions on team formation. One participant (P2) preferred student-formed teams, but it was a source of anxiety, saying,
    \textit{``Group work is super overwhelming. Especially because I'm also shy. \textbf{I worry about not getting picked by anyone or not getting along with the people I'm  paired with}.''}
However, another participant (P1), when reflecting on a past experience about group work, described how they had a positive experience when they were able to choose to work with people who had similar interests and shared identities,
    \textit{``Her and I were really vibing, and she is also \textbf{very much neurodivergent and very much passionate about the same subject}.''}

\subsubsection{Team Size}
ND students felt the most comfortable with teams of 3--5 students, then teams of 2 students, and lastly teams of 6 or more students.
NT students felt the most comfortable in teams of 2 students. 
All of our interview participants expressed that they preferred group sizes which did not introduce logistical issues, such as coordinating with a large group. They also shared that some team sizes introduce uncomfortable social dynamics,
such as working with only one person. P1 said,
    \textit{``I think that the 3-4 range of people kind of lets everybody have a chance to participate... \textbf{in larger groups, I think people are sometimes just trying to talk over each other} and then in smaller groups like only 2 people... \textbf{I feel like I would be more uncomfortable with only working one person.}''}

\subsubsection{Frequency of Collaboration}
ND students felt the most comfortable with teams that met on a recurring basis or teams that worked together one time
over a semester-long team.
NT students had the same order of preference, however ND students felt substantially less comfortable with semester-long teams. 50\% of ND students reported feeling some level of discomfort compared to 22\% of NT students.
In our interviews, participants (P1, P2, P4) described how recurring teams may give neurodivergent students the opportunity to develop a stronger understanding of how the team works together. P2 said, 
    \textit{``I tend to get more comfortable towards the end, because it \textbf{takes me a while to understand what the group dynamics are like}.''}
Additionally, consistent teams may reduce anxiety surrounding collaborative learning. P4 said, ``\textit{\textbf{It was kind of comforting to know that I didn't have to worry about the groups changing} after they were fully assigned by the professor.}''
\subsubsection{Student Roles}
\label{sec:roles}
ND students felt the most comfortable when students had roles that had been chosen either from a set of instructor-provided roles, independently chosen roles, or instructor-assigned roles over having no specific roles.
NT students felt most comfortable when roles were chosen independently, then chosen from an instructor-provided set, then having no specific roles, then having instructor-assigned roles.
In our interviews, multiple participants felt that students having specific roles was helpful in facilitating collaboration and reduces anxiety about managing the workload (P2, P3, P4). Specifically, P3 said, 
    \textit{``Role assignment would be helpful to put me to a specific task because if \textbf{I don't have a task I feel like I have to check on everything}... If everyone doesn't have a specific task, I feel like there are some things that might be an oversight.''}

\subsection{RQ2: Pedagogical \& Environmental Factors}
\label{sec:other-impacts}
Other aspects of CAL may impact students' comfort: \textit{assignment structure, instructor involvement, and the environment~\cite{shekhar2020negative,gordy2020multi,greer2019effects}}. We show these results using the distribution of participants who responded ``Somewhat comfortable'' to ``Very comfortable.''

\subsubsection{Structure Provided for Assignments}
\label{sec:structure}
ND and NT students indicated similar levels of comfort with the amount of structure provided for assignments. They felt most comfortable with fully structured assignments (ND: 87.5\%; NT: 65\%), then assignments with outlines or guidelines provided (ND: 66.67\%; NT: 60\%), and lastly assignments with no structure provided (ND: 16.67\%; NT: 35\%). ND students tended to be more comfortable with some level of structure provided, and less comfortable with unstructured assignments. In our interviews, participants (P2, P3, P4) expressed that they felt uncomfortable when assignment specifications allowed room for error or provided ambiguous instructions. P4, while reflecting on the scripted activities their instructor has them do, said,
    \textit{``The structure is very helpful compared to activities with less structure because \textbf{there's a lot of room for error in both coding and also interpreting what I'm supposed to do.}''}
Another participant, P1, expressed that unclear expectations not only made them uncomfortable, but made them feel like they had a lower chance of succeeding,
    \textit{``I would very much appreciate, like, tell me what I need to have a good grade. \textbf{I cannot read your mind}.''}

\subsubsection{Instructor Involvement}
ND and NT students indicated similar comfort levels with varying levels of instructor involvement. They felt the most comfortable with some instructor involvement (ND: 62.5\%; NT: 60\%), then high instructor involvement (ND: 58.33\%; NT: 57.89\%), and lastly fully student-directed (ND: 25\%; NT: 30\%).  
All of our interview participants described how some instructor involvement would make them feel more comfortable, knowing that they were meeting the instructor's requirements, on the right track, and able to get assistance beyond their team members. P2 said,
    \textit{``I would prefer the professor or the instructor to be more involved, because \textbf{that give [us] more feedback and help us guide through what expectations are at each step more concretely}.''}

\subsubsection{Location}
Both ND and NT students expressed that they were more comfortable with in-person collaboration (ND: 57.17\%; NT: 80\%) over online collaboration (ND: 29.17\%; NT: 45\%). In our interviews, some participants (P2, P4) described feeling more comfortable with in-person because they felt like there was less guesswork in understanding their teammates. P2 said,
    \textit{``Between Zoom and in-person, I would probably prefer in-person... I sometimes feel \textbf{it's harder to kind of gauge what they're actually thinking.}''}

\subsection{RQ3: Coping Methods \& Recommendations}
\label{sec:interview-results}

\subsubsection{Coping Methods}
In our interviews, a common coping method that came up with participants was self-assignment to the same types of roles when working in teams. Some of our participants (P2, P3, P4) said that they often took on more clerical or managerial roles, describing how it allowed them to have more control over the way work was conducted, structured, and formatted. P2 talked about how they preferred to take a more managerial role, saying,
    \textit{``I don't like leading it, but I do kind of, like, end up \textbf{directing how we want to it and structure the overview of the project}.''}
On the other hand, P1 described being the ``leader'' of their group, especially when working on a project they had a lot of knowledge and interest in,
    \textit{``I would characterize my role in the group as like, \textbf{the leader, the captain. It's like, I know way too much about this subject. I can help}.''}
In our survey results (see Section~\ref{sec:roles}), ND participants described being the most comfortable with student-selected roles, either from a set provided by the instructor or independently chosen by students. This may be because it provides students with the opportunity to choose roles that align with what tasks they find themselves most comfortable with. P4 described their preference for the role they often chose, 
    \textit{``Usually, I like doing [roles] where you just kind of write everything down that the group is doing and submit the assignment, because that one, \textbf{it's like, more passive}.''}
This may be a result of how they experience their neurodiversity, P4 discussed how they find interacting with their teammates difficult,
    \textit{``Trying to participate in groups without being overpowering and talking over everyone or the opposite and staying back and doing nothing is \textbf{a balance I can sometimes not find easily}.''}

Two of our participants (P2, P3) shared that they have a lot of expectations and preferences for the way work is done. 
P2 went on to describe how this can cause conflict within their groups, especially when their teammates do not understand why they may need a more structured approach to collaborative work. They found that being very explicit with their access needs~\cite{invalid2017skin} (i.e., the things people need to fully participate in a space or activity) was helpful, stating,
    \textit{``[Other students] don't really know that I struggle with some things, \textbf{they would consider me as just another person who is neurotypical}, and they wouldn't really know that I have some expectations.''}
Beyond wanting explicit guidelines and expectations for assignments from instructors, as found in our survey results (see Section~\ref{sec:structure}), participants highlighted the need for explicit expectations from their teammates as well.

Participants also described how they cultivated a sense of belonging within their teams by searching for people who have similar interests and identities. Two participants (P1, P2) expressed that they liked the experience of working with other ND students. P2 said,
    \textit{``If I do have someone who's also kind of like me, then I think we would get along well... 
    \textbf{If you're both neurodivergent and have similar thought processes or approaches to things}.''}

\subsubsection{When Coping Methods Fail}
These coping methods are not perfect, and multiple participants (P1, P2, P3), described scenarios in which the strategies they have used before did not work to make them feel more comfortable or were not applicable. In many cases, the failure of their coping mechanisms resulted in avoidance of socialization in their courses and CAL which requires a lot of public presentations. P2 described how they have a sense of anxiety about public perception, and they would rather drop the class entirely than participate in those assignments,
    \textit{``\textbf{I generally don't like being perceived}... so even when I do take classes [that require frequent presentations to the entire class], I tend to sit in a corner.''}
Beyond discomfort in collaborative assignments, one participant (P1) expressed that they have a lot of anxiety surrounding the idea of collaboration in their computing courses. They expressed that their neurodiversity, gender, and sexuality factored into a reduced sense of belonging and feeling ostracized,
    \textit{``I'm non-binary, I'm queer... 
    I feel like I get less respected, and I feel like... \textbf{I'm the odd one out}.''}

\subsubsection{Recommendations for Improved Accessibility}

In some of our interviews, participants (P1, P3) described ways in which they felt like they were ``\textit{fighting against the architecture}'' (P1) of computing spaces. In this context, we use the term architecture broadly. In the literal sense the architecture of buildings and classrooms may negatively impact ND students’ experiences. In an abstract sense, the architecture or structure of computing curricula which has been developed primarily with neurotypical students in mind may negatively impact ND students’ experiences. P1 described how they felt alienated from their classmates based solely on the room in which they were in, ``\textit{\textbf{We're fighting against this space we have}, especially in lecture halls, where it's all lined up, where you don't really meet your neighbor.}''
Beyond the physical surroundings of their computing classrooms, P1 stressed how they felt that the curricula of their computing classrooms was not structured to support social interactions and developing relationships with their peers, and they suggested that low-stakes collaborative activities could be introduced earlier and more frequently.
Both participants highlighted the need for change within computing courses to provide learning environments which fosters collaboration.

Participants (P1, P2, P3) emphasized the importance of providing explicit requirements and expectations. P2 said,
    \textit{``\textbf{Assignments should have clear guidelines as to what is expected}... What is expected at the end of each iteration and what is considered `done.'''}
Participants (P2, P3) also expressed that introducing consistent feedback would help keep them and their teammates on track and mitigate feelings of being ``\textit{left behind}'' (P3).
Lastly, participants described ways in which instructors have made CAL more engaging. Participants (P2, P3) discussed how projects should be able to incorporate personal interests or real-world applications as much as possible. P2 said,
   \textit{``Projects where I can see real-world applications especially triggered my hyper-focus. \textbf{I learned more in 10 days than 6 weeks of generic assignments.}''}
While participants described ways in which CAL could be made more accessible, nearly all participants (P2, P3, P4) found CAL beneficial for their learning.

\section{Discussion} 

Autistic students often experience difficulties with socialization~\cite{clouder2020neurodiversity,zolyomi2018values}. To some, this might indicate that making CAL more accessible for autistic students requires alternative assessments that do not require collaboration. Our findings do not indicate that. In our interviews, participants believed collaborative learning was beneficial, even with its challenges. They did not indicate a desire to remove it from their courses. We recommend that researchers and practitioners incorporate collaboration earlier and more often with careful considerations. 
For example, researchers have proposed techniques to support autistic computing students in developing teamwork skills and communication self-efficacy in a video game coding camp~\cite{begel2021remote}. Instructors can also incorporate the idea of ``social translucence'' as presented by Zolyomi et al.~\cite{zolyomi2018values}, which provide guidance about supporting neurodiverse teams with strategies to promote visibility, accountability, awareness, and identity.

Autistic students or students with ADHD often struggle with balancing workloads and experience difficulties with hyper-focus and executive functioning~\cite{clouder2020neurodiversity,zolyomi2018values} and autistic students may struggle with unpredictable environments~\cite{gin2020active,pilotte2016autism}.
Building in guardrails to assignments that promote consistency and predictability (e.g., team contracts, regular schedules, consistent feedback loops) may help to develop a more comfortable environment for autistic students and students with ADHD~\cite{salvatore2024not}.
Additionally, autistic students often struggle with deciphering unclear and ambiguous instructions and expectations~\cite{egan2005students,pilotte2016autism,sharmin2024towards,stuurman2019autism}. Our survey and interview results provide evidence to support recommendations from previous position papers recommending instructors develop explicit and unambiguous instructions and expectations for assignments~\cite{egan2005students,sharmin2024towards,stuurman2019autism}. Lastly, our results indicate that students value their ability to exercise agency over who they work with, their roles within teams, and the ability to pursue topics which are interesting to them. Existing research has begun to explore the development and use of personally relevant pedagogy for minoritized students and on large scales~\cite{bernstein2024like, mckinney2024iterative}.

We caution against interpreting our results as applicable to all ND students. Neurodiversity, and disability more broadly, is a spectrum experienced differently by each individual~\cite{lutz2005disability}. Often, neurodiversity is not an isolated identity. It is an intersectional experience, intertwined with factors like race, sexuality, and gender~\cite{runa2023student, bonnette2025need, tadimalla2025connecting}, so cultivating an environment which feels safe for people of all identities is vital to maintaining and serving a diverse student body within computing education~\cite{calabrese2020beyond}. 
In our work, we highlight several factors which should be considered in designing equitable CAL techniques: team dynamics, environment (e.g., classroom layout, collaboration location), assignment structure, and instructor involvement. This is not an exhaustive list. Additionally, some of our results highlight where preferences of ND and NT students align. These points of alignment could inform curricula development that benefits more than just ND students~\cite{murray2023human}. Researchers should continue to work with ND students to understand their preferences, experiences, and needs, while practitioners work directly with students in their courses. We also identified self-disclosure of ND identities and their access needs as a coping method to adapt to CAL. While self-disclosure can be potentially effective in receiving accommodations, it can be uncomfortable, emotionally taxing, and expose students to bias and stigma~\cite{au2025gonna,newman2025disclosure}. Anonymous disclosure of access needs may facilitate ND students in receiving accommodations while reducing the burden of self-disclosure.

\section{Limitations \& Future Work}
Our two studies focus on the preferences and experiences of computing students with two ND identities with CAL. Our sample sizes are small and geographically limited. Recruiting participants for accessibility research is challenging~\cite{sears2011representing}, and studies with ND and disabled participants tend to have lower participant numbers~\cite{mack2021we,figueira2026where}. Our findings are not meant to be generalized to the broader ND computing student population, as neurodiversity is a highly individual experience. We present preliminary insights into some preferences and experiences of ND computing students, specifically those on the autism spectrum and with ADHD. 
Future work should include more ND students to validate the trends we observed and discover additional insights. Future work should also investigate the impacts of CAL with more ND student populations, such as 
students with dyslexia, learning disabilities, or mental health conditions.

\section{Conclusion}
In this paper, we described the results of a survey conducted with 44 computing students (24 ND, 20 NT) and interviews with 4 ND computing students. While our results are preliminary, they highlight various interesting trends and open avenues for future work. In our survey, ND students provided insights into what aspects of CAL make them feel more comfortable. Our interviews illuminated some coping mechanisms ND students use to adapt to CAL and their recommendations to make it more accessible. Our work adds to a growing body of research by providing empirical results from ND computing students and highlighting promising directions for researchers and practitioners to design CAL techniques that support a more equitable learning environment for students.

\begin{acks}
This work was supported by the Temple University College of Science and Technology Summer Undergraduate Research Experience.
\end{acks}

\bibliographystyle{ACM-Reference-Format}
\bibliography{sample-base}

\end{document}